\documentclass[oneside,english,numreferences]{kluwer}
\usepackage[T1]{fontenc}
\usepackage[latin1]{inputenc}
\usepackage{graphicx}
\usepackage{amssymb}

\makeatletter

\newcommand{\noun}[1]{\textsc{#1}}
\usepackage{graphicx}

\usepackage{klucite}

\newcommand{\comment}[1]{}
\newcommand{\etal}[1]{, #1}

\usepackage{babel}
\makeatother
\begin{document}

\newcommand{\bra}[1]{\langle#1 |}

\newcommand{\ket}[1]{| #1 \rangle}

\newcommand{\expect}[1]{{\left\langle #1 \right\rangle }}

\newcommand{\spup}{\ket{\!\uparrow}}

\newcommand{\spdown}{\ket{\!\downarrow}}

\newcommand{\spupbra}{\bra{\uparrow\!}}

\newcommand{\spdownbra}{\bra{\downarrow\!}}

\newcommand{\spupup}{\ket{\!\uparrow\uparrow}}

\newcommand{\spupdown}{\left.\ket{\!\uparrow\downarrow}\right.}

\newcommand{\spdownup}{\ket{\!\downarrow\uparrow}}

\newcommand{\spdowndown}{\ket{\!\downarrow\downarrow}}

\newcommand{\sqrtSwap}{U_{\mathrm{sw}}^{1/2}}

\newcommand{\efficiencyMeas}{e}

\newcommand{\measOutcome}[1]{A_{#1}}

\newcommand{\POVMop}[1]{E_{\measOutcome{#1}}}

\newcommand{\binomProb}[1]{p_{#1}}

\newcommand{\gammaOut}{\gamma_{\mathrm{out}}}

\newcommand{\gammaIn}{\gamma_{\mathrm{in}}}

\newcommand{\timeQPCmeas}{t_{\mathrm{m}}}

\begin{opening}

\title{Controlling Spin Qubits in Quantum Dots}

\author{Hans-Andreas \surname{Engel}}

\institute{Department of Physics and Astronomy, University of Basel, Klingelbergstrasse 82, CH-4056 Basel, Switzerland}

\author{L.P. \surname{Kouwenhoven}}

\institute{Department of NanoScience and ERATO Mesoscopic Correlation Project, Delft University of Technology, P.O. Box 5046, 2600 GA Delft, The Netherlands}

\author{Daniel \surname{Loss}}

\institute{Department of Physics and Astronomy, University of Basel, Klingelbergstrasse 82, CH-4056 Basel, Switzerland}

\author{C.M. \surname{Marcus}}

\institute{Department of Physics, Harvard University, Cambridge, Massachusetts 02138, USA}

\begin{abstract}
We review progress on the spintronics proposal for quantum computing
where the quantum bits (qubits) are implemented with electron spins.
We calculate the exchange interaction of coupled quantum dots and
present experiments, where the exchange coupling is measured via transport.
Then, experiments on single spins on dots are described, where long
spin relaxation times, on the order of a millisecond, are observed.
We consider spin-orbit interaction as sources of spin decoherence
and find theoretically that also long decoherence times are expected.
Further, we describe the concept of spin filtering using quantum dots
and show data of successful experiments. We also show an implementation
of a read out scheme for spin qubits and define how qubits can be
measured with high precision. Then, we propose new experiments, where
the spin decoherence time and the Rabi oscillations of single electrons
can be measured via charge transport through quantum dots. Finally,
all these achievements have promising applications both in conventional
and quantum information processing.
\end{abstract}

%\abbreviations{\abbrev{...}{...}; \abbrev{...}{...}}

\end{opening}

\section{Introduction}

\label{secIntro}

The spin degree of freedom promises many applications in electronics~\cite{PrinzPT,Wolf,spintronicsBook}.
Prominent experiments have shown injection of spin-polarized currents
into semiconductor material~\cite{Fiederling,Ohno}, long spin dephasing
times in semiconductors (approaching microseconds) \cite{Kikkawa},
ultrafast coherent spin manipulation \cite{guptascience}, as well
as phase-coherent spin transport over distances of up to $100\,\mu{\textrm{m}}$~\cite{Kikkawa}.
Irrespective of spin, the charge of the electrons can be used to control
single electrons by confining them in quantum dot structures, which
leads to striking effects in the Coulomb blockade regime \cite{Kouwenhoven}.
The Loss and DiVincenzo proposal~\cite{Loss97} combines these two
fields of research and uses the spin of electrons confined on quantum
dots as spin qubits for quantum computation. This proposal comprises
two-qubit quantum gates relying on the exchange interaction of coupled
quantum dots and comprises spin-to-charge conversion for efficient
read-out schemes, satisfying all theoretical requirements for quantum
computing. This quantum computer proposal, based on exchange interaction,
can be mapped from electron spins on dots to nuclear spins of P atoms
in Si, as shown by Kane \cite{Kane} (see article in this issue).

The spin qubit proposal~\cite{Loss97} addresses the central issues
for building a quantum computer. However, for a concrete implementation
of spin qubits, a more detailed theoretical and experimental understanding
of spins on quantum dots is required. This demand has led to many
new theoretical and experimental investigations on quantum dots, which
also address interesting aspects of physics on their own. In this
article we will review some of these recent results.

\subsection{Quantum Dots}

\label{ssecQD}

\begin{figure}
 \centerline{\includegraphics[%
  width=11.5cm]{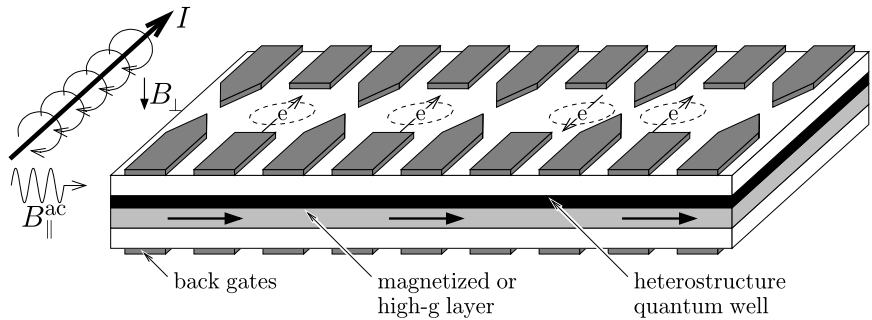}}

\caption{An array of quantum dots (circles) is defined by gate electrodes
(dark gray) which confine the electrons. For spin manipulations, electrons
can be moved by changing the gate voltage, pushing the electron wave
function into the magnetized or high-$g$ layer, allowing for spatially
varying Zeeman splittings. Alternatively, local magnetic fields can
be achieved by a current-carrying wire (indicated on the left of the
dot array). Then, the electron in each dot is subject to a distinct
Zeeman splitting. This can be used for one-qubit gates, since only
relative spin rotations are sufficient. Further, the spins can be
addressed individually with ESR pulses of an oscillating in-plane
magnetic field which is in resonance with a particular Zeeman splitting.
These mechanisms allow single-spin rotations in different spatial
directions. For gate operations on two qubit spins, their exchange
coupling can be controlled by lowering the tunnel barrier between
the dots, see Sec.~\ref{sec2bit}. Here, the two rightmost dots are
drawn schematically as tunnel-coupled. Note that only electrical switching
is required to control spin dynamics and quantum computation with
such a device. }

\label{figArray}
\end{figure}

In this article we consider semiconductor quantum dots. These are
structures where charge carriers are confined in all three spatial
dimensions. The dot size, typically between $10\:{\textrm{nm}}$ and
$1\:\mu\mathrm{m}$ \cite{Kouwenhoven}, is on the order of the Fermi
wavelength in the host material. The confinement of the quantum dots
is usually achieved by electrical gating of a two-dimensional electron
gas (2DEG), possibly combined with etching techniques, see Figs.~\ref{figArray},
\ref{fig:doubleDot}(a), and \ref{fig:filterLarge}(b). Small dots
have charging energies in the meV range, resulting in quantization
of charge on the dot (Coulomb blockade). This allows precise control
of the number of electrons and of the spin ground state on the dot.
Such a control of the number of electrons in the conduction band of
a quantum dot (starting from zero) has been achieved with GaAs heterostructures,
e.g., for vertical dots \cite{tarucha} and lateral dots \cite{ddSmallLK,SachrajdaSmallDot}.
Quantum dots have various tunable parameters. These include geometry,
energy spectrum, coupling between dots, etc., which open up many possibilities
by providing a versatile system for manipulation of electronic states,
in particular the spin state. Further, the electronic dot-orbitals
are highly sensitive to external magnetic and electric fields \cite{Kouwenhoven,tarucha},
since the magnetic length corresponding to fields of $B\approx1\,{\textrm{T}}$
is comparable to typical dot sizes. In coupled quantum dots, Coulomb
blockade effects~\cite{waugh}, tunneling between neighboring dots~\cite{Kouwenhoven,waugh,ddReviewVDWiel},
and magnetization~\cite{oosterkamp} have been observed as well as
the formation of a delocalized single-particle state~\cite{blick,ddLargeTunnelCM}
and coherent charge oscillations~\cite{FujisawaT2dd}. 

\subsection{Quantum Computing with Spin Qubits}

\label{ssecSpinQubits}The interest in quantum computing~\cite{Steane,NielsenChuang}
derives from the hope to outperform classical computers using new
quantum algorithms. These algorithms make use of the quantum computer's
abilities to exist in a quantum superpositions of its {}``binary''
basis states $|0\cdots00\rangle$, $|0\cdots01\rangle$, $|0\cdots10\rangle$,...,
and to perform unitary time evolutions $U|\Psi_{\mathrm{in}}\rangle=|\Psi_{\mathrm{out}}\rangle$
for computation. The basis states can be realized by concatenating
several quantum bits (qubits) which are states in the Hilbert space
spanned by $\ket{0}$ and $\ket{1}$. A natural candidate for the
qubit is the electron spin because every spin $\frac{1}{2}$ encodes
exactly one qubit. Such spin qubits on quantum dots are good candidates
for realizing a quantum computer~\cite{Loss97}. We consider the
five criteria of DiVincenzo's checklist \cite{DiVincenzoListFdp}
which must \emph{all} be satisfied for any physical implementation
of a quantum computer. We briefly discuss that these criteria are
satisfied for spins qubits \cite{Loss97,FdP}. These criteria provides
us with a good starting point for going into the details of concrete
parts of the actual implementation of spin qubits. In the following
sections we then show where specific theories and current experiments
give new insight into the realization of spin qubits.

i) \emph{A scalable system with well characterized qubits} is required.
To speed up calculations using a quantum computer, one needs a large
number of qubits, i.e., on the order of $10^{5}$.  This requirement
is achievable for spin qubits, since producing arrays of quantum dots
is feasible with state-of-the-art techniques for defining nanostructures
in semiconductors. Further, the electron's spin~$\frac{1}{2}$ provides
a natural qubit, setting $\ket{0}\equiv\spup$ and $\ket{1}\equiv\spdown$.

ii) \emph{The state of the qubits must be initialized} to a known
value at the beginning of a computation. To initialize spin qubits,
one can apply a large magnetic field $g\mu_{B}B\gg kT$ that allows
them to relax to the thermal ground state. Alternatively, one can
inject polarized electrons into the dot by using spin-polarizing materials~\cite{Fiederling,Ohno}
or by using a spin filter~\cite{Recher} which we describe in Sec.~\ref{ssecSpinFilter}. 

iii) \emph{Long decoherence times, much longer than the gate operation
time}, is the most difficult criterion to satisfy for many quantum
computer proposals. Here, the current knowledge about the spin qubits
is very promising. Gate operation times well below one ns are in principle
feasible \cite{FdP}. Using theoretical estimates and experimental
data on spin flip times, the expected decoherence times can reach
ms, see Sec.~\ref{secRelax}. Thus, the decoherence times could be
eight orders of magnitude larger than the gate operation times. 

iv) With \emph{a universal set of quantum gates}, any quantum algorithm
can be implemented by controlling a particular unitary evolution of
the qubits. It is sufficient to have single-qubit gates and a universal
two-qubit gate (e.g., \textsc{xor} or square root of \textsc{swap}).
Single qubit gates can be produced by controlling the local magnetic
field, the local $g$ factor (or $g$ tensor), or local Overhauser
field, which, e.g., can be achieved with a semiconductor heterostructure
and electrical gating \cite{FdP,SalisGfactor,kato}, see Fig.~\ref{figArray}.
To build two-qubit gates, one can use the exchange interaction which
arises when two neighboring dots are tunnel coupled, which can again
be controlled via gate voltages \cite{Loss97,BLD}. We describe the
exchange interaction of coupled dots in Sec.~\ref{sec2bit}.

v) \emph{Qubit read out} determines the result at the end of the computation
by measuring specific qubits. There are several proposals for measuring
the spin in quantum dots, most of them rely on transferring the information
from the spin to the charge state \cite{Loss97}, e.g., by making
use of the Pauli principle \cite{Recher,ELesr,ELreadout}, via the
spin-orbit interaction \cite{LevitovRashba}, or by making use of
the Zeeman splitting \cite{ELreadout}. We discuss concrete read-out
schemes for spin qubits in Sec.~\ref{secSpinMeas} and address experiments~\cite{ssReadOutDelft}
where single-shot read out has been achieved.

\section{Two Coupled Quantum Dots as Quantum Gates}

\label{sec2bit} 

We now consider a pair of spin qubits which are coupled by the exchange
interaction, which results from the combination of the Coulomb interaction
and the Pauli exclusion principle. Two electrons in coupled quantum
dots and in the absence of a magnetic field have a spin-singlet ground
state, while the first excited state in the presence of sufficiently
strong Coulomb repulsion is a spin triplet. Higher excited states
are separated from these two lowest states by an energy gap, given
either by the Coulomb repulsion or the single-particle confinement.
The low-energy dynamics of such a system is described by the effective
Heisenberg spin Hamiltonian, \begin{equation}
H_{\textrm{s}}(t)=J(t)\,\,{\textbf{S}}_{1}\cdot{\textbf{S}}_{2},\label{Heisenberg}\end{equation}
 where $J(t)$ describes the exchange coupling between the two spins
${\textbf{S}}_{1}$ and ${\textbf{S}}_{2}$ and is given by the energy
difference between the triplet and the singlet, $J=E_{T_{0}}-E_{S}$.
After a pulse of $J(t)$ with $\int_{0}^{\tau_{s}}dtJ(t)/\hbar=\pi$
(mod $2\pi$), the time evolution $U(t)=T\exp(i\int_{0}^{t}H_{\textrm{s}}(\tau)d\tau/\hbar)$
corresponds to the \noun{swap} operator $U_{\textrm{sw}}$, whose
application leads to an interchange of the two spin states. While
$U_{\textrm{sw}}$ is not sufficient for quantum computation, any
of its square roots, say $\sqrtSwap\ket{\phi\chi}=(\ket{\phi\chi}+i\ket{\chi\phi})/(1+i)$,
turns out to be a \emph{universal} quantum gate. It can be used, together
with single-qubit rotations, to assemble any quantum algorithm~\cite{Loss97}. 

\begin{figure}
\includegraphics[%
  width=70mm,
  keepaspectratio]{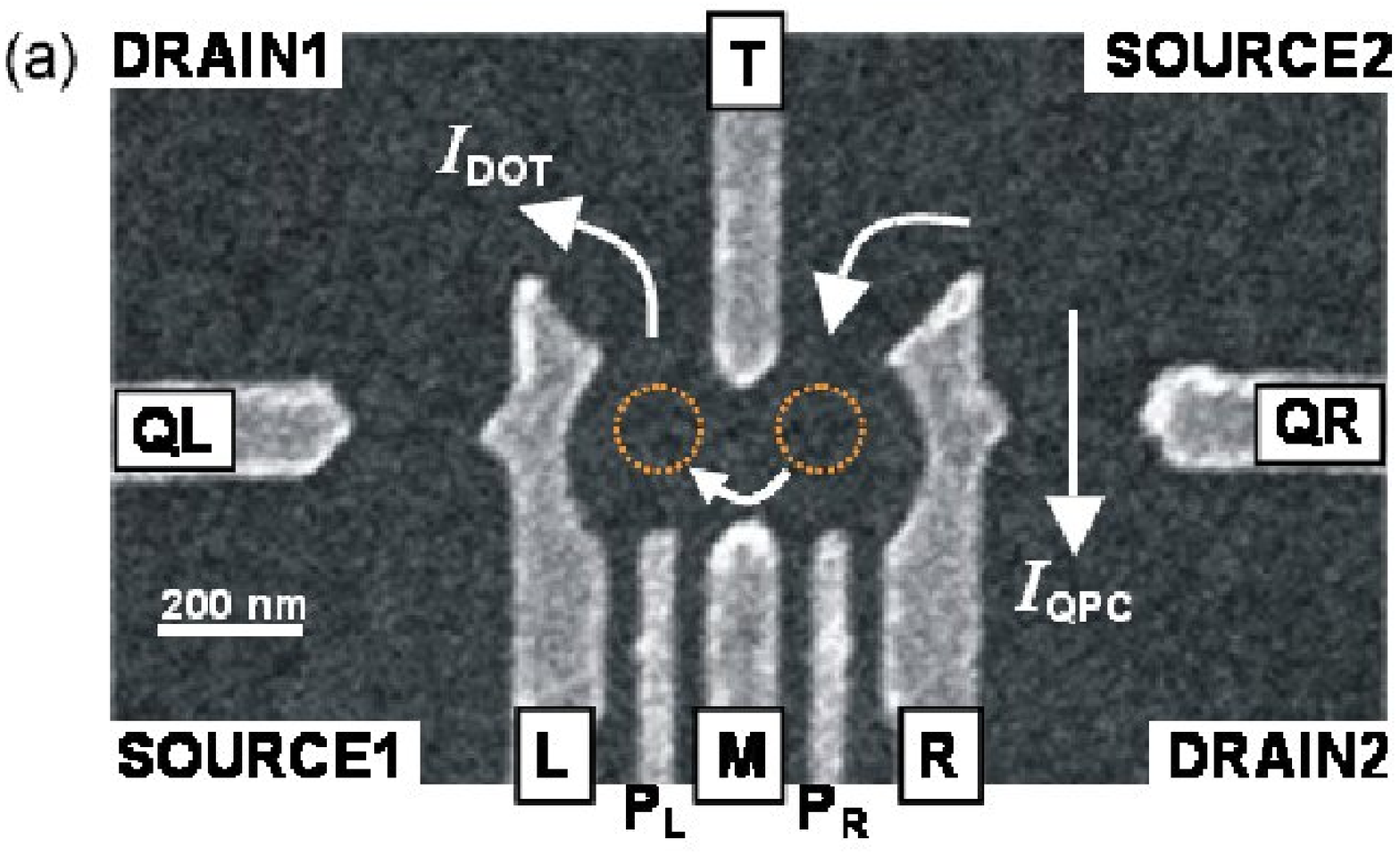}~~\includegraphics[%
  width=50mm]{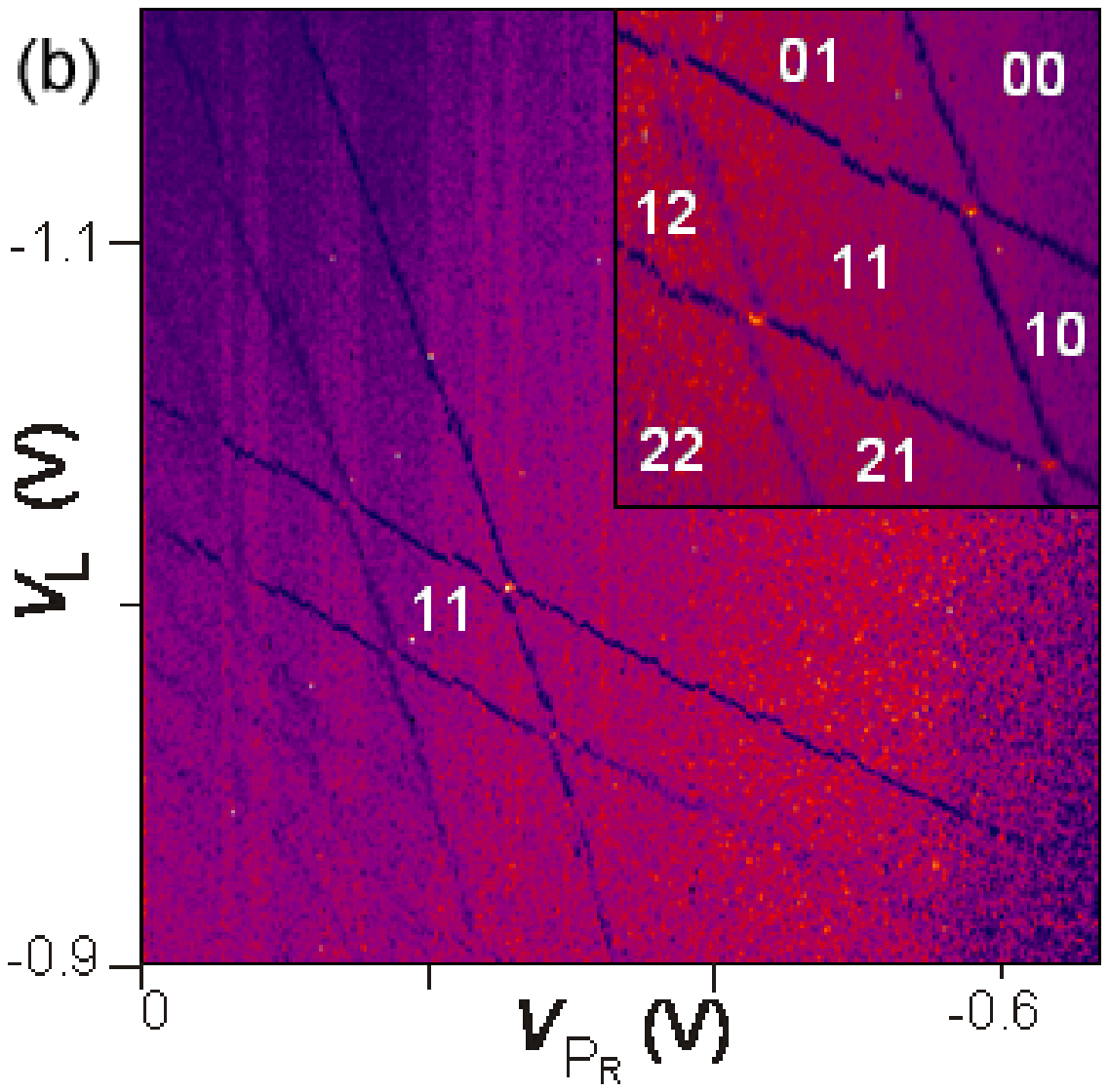}

\caption{(a) Double dot structure with a single electron in each dot, shown
as scanning electron micrograph of the metallic surface gates \cite{ddSmallLK}.
The circles indicate the two quantum dots and the arrows show the
possible current paths. A bias voltage, $V_{\mathrm{DOT}}$ , can
be applied between source 2 and drain 1, leading to current through
the dots. A bias voltage, $V_{\mathrm{SD}i}$ between source and drain
$i=1,2$ yields a current, $I_{\mathrm{QPC}}$, through the corresponding
QPC. (b) Charge stability diagram (honeycomb) \cite{ddReviewVDWiel}
of the double quantum dot, measured with QPC-R \cite{ddSmallLK}.
A modulation (0.3 mV at 17.77 Hz) is applied to gate L, and $dI_{\mathrm{QPC}}/dV_{\mathrm{L}}$
is measured with a lock-in amplifier and plotted versus $V_{\mathrm{L}}$
and $V_{\mathrm{PR}}$. The bias voltages are $V_{\mathrm{SD}2}=100\:µ\mathrm{V}$
and $V_{\mathrm{DOT}}=V_{\mathrm{SD1}}=0$. The inset shows a magnification
of the honeycomb pattern for the first few electrons in the double
dot. The labels {}``$n_{\mathrm{L}}n_{\mathrm{R}}$'' indicate the
number of electrons in the left and right dot, and the double dot
is completely empty in the region {}``00.'' }

\label{fig:doubleDot}
\end{figure}

We consider a system of two coupled quantum dots in a 2DEG, containing
one (excess) electron each, see Fig.~\ref{fig:doubleDot}(a). The
dots are arranged in a plane such that the electrons can tunnel between
the dots, leading to an exchange interaction $J$ between the two
spins, which we now calculate. We model this system of coupled dots
with the Hamiltonian $H=\sum_{i=1,2}h_{i}+C+H_{\textrm{Z}}=H_{\textrm{orb}}+H_{\textrm{Z}}$.
The single-electron dynamics in the 2DEG ($xy$-plane) is defined
with the Hamiltonian $h_{i}$, containing the quartic confinement
potential\begin{equation}
V(x,y)=\frac{m\omega_{0}^{2}}{2}\left[\frac{1}{4a^{2}}\left(x^{2}-a^{2}\right)^{2}+y^{2}\right],\label{eqnV}\end{equation}
 with inter-dot distance $2a$, effective Bohr radius $a_{\textrm{B}}=\sqrt{\hbar/m\omega_{0}}$,
and effective mass $m$. Separated dots ($a\gg a_{\textrm{B}}$) are
thus modeled as two harmonic wells with frequency $\omega_{0}$, consistent
with experiments where the low-energy spectrum of single dots indicates
a parabolic confinement~\cite{tarucha}. A magnetic field ${\textbf{B}}=(0,0,B)$
is applied along the $z$-axis, which couples to the electron spins
through the Zeeman interaction $H_{\textrm{Z}}$ and to the charges
through the vector potential $\textbf{A}(\textbf{r})=\frac{B}{2}(-y,x,0)$.
In almost depleted regions, like few-electron quantum dots, the screening
length $\lambda$ can be expected to be much larger than the screening
length in bulk 2DEG regions (where it is $40\,\textrm{nm}$ for GaAs).
Thus, for small quantum dots, say $\lambda\gg2a\approx40\,\textrm{nm}$,
we consider the bare Coulomb interaction $C=e^{2}/\kappa|\textbf{r}_{1}-\textbf{r}_{2}|$,
where $\kappa$ is the static dielectric constant. 

\begin{figure}
 \centerline{\includegraphics[%
  width=5cm]{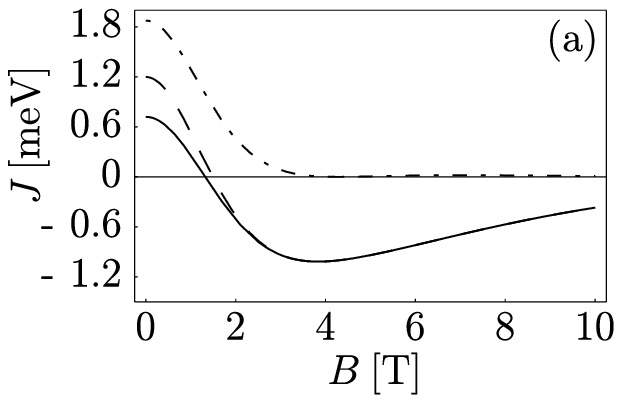}~\includegraphics[%
  width=40mm]{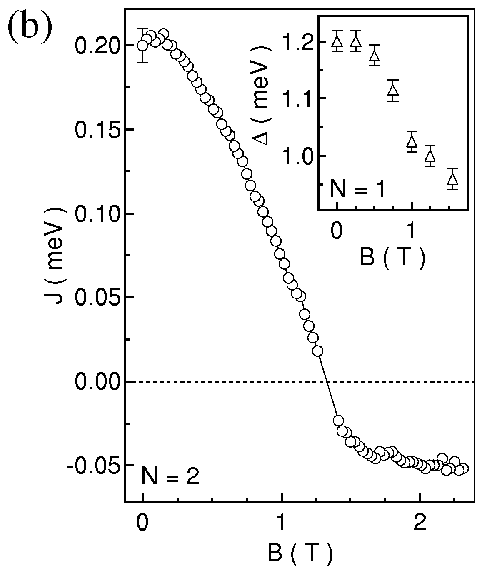}}

\caption{The exchange coupling $J$ (solid line) for GaAs quantum dots a function
of the magnetic field $B$. (a) Theoretical prediction {[}Eq. (\ref{J}){]}
for a double dot with confinement energy $\hbar\omega=3\,{\textrm{meV}}$,
inter-dot distance $d=a/a_{\textrm{B}}=0.7$, and $c=2.42$~\cite{BLD}.
For comparison, the short-range Hubbard result $J=4t^{2}/U$ (dashed-dotted
line) and the extended Hubbard result $J=4t^{2}/U+V$ (dashed line)
are plotted. (b) Experimentally observed exchange coupling $J$ via
transport measurements~\cite{exchangeZumbuhl}. Although a single
dot structure was used, the measurements show double dot features,
indicating that a double dot is formed within the structure. The dependence
on magnetic field $B$ is in agreement with the theoretical predictions,
in particular, $J$ can be tuned through zero near $B=1.3\:\mathrm{T}$.}

\label{Jplots}
\end{figure}

Now we consider only the two lowest orbital eigenstates of $H_{\textrm{orb}}$,
leaving us with one symmetric (spin singlet) and one antisymmetric
(spin triplet) orbital state. The spin state for the singlet is $\ket{S}=(\spupdown-\spdownup)/\sqrt{2}$,
while the triplet spin states are $\ket{T_{0}}=(\spupdown+\spdownup)/\sqrt{2}$,
$\ket{T_{+}}$=$\spupup$, and $\ket{T_{-}}$=$\spdowndown$. For
$kT\ll\hbar\omega_{0}$, higher-lying states are frozen out and $H_{\textrm{orb}}$
can be replaced by the effective Heisenberg spin Hamiltonian {[}Eq.~(\ref{Heisenberg}){]}.
To calculate the triplet and singlet energies, we use the analogy
between atoms and quantum dots and make use of variational methods
similar to the ones in molecular physics. Using the Heitler-London
ansatz with the ground-state single-dot orbitals, we find~\cite{BLD},
\begin{eqnarray}
J & = & \frac{\hbar\omega_{0}}{\sinh\left(2d^{2}\,\frac{2b-1}{b}\right)}\Bigg\{\frac{3}{4b}\left(1+bd^{2}\right)\nonumber \\
 &  & +c\sqrt{b}\left[e^{-bd^{2}}\, I_{0}\left(bd^{2}\right)-e^{d^{2}(b-1/b)}\, I_{0}\left(d^{2}\,(b-1/b)\right)\right]\Bigg\},\label{J}\end{eqnarray}
 with zeroth order Bessel function $I_{0}$, dimensionless distance
$d=a/a_{\textrm{B}}$ between the dots, magnetic compression factor
$b=\sqrt{1+\omega_{L}^{2}/\omega_{0}^{2}}$, and Larmor frequency
$\omega_{L}=eB/2mc$. In Eq.~(\ref{J}), the first term arises from
the confinement potential, while the terms proportional to the parameter
$c=\sqrt{\pi/2}(e^{2}/\kappa a_{\textrm{B}})/\hbar\omega_{0}$ result
from the Coulomb interaction $C$; the exchange term is recognized
by its negative sign. We are mainly interested in the weak coupling
limit $|J/\hbar\omega_{0}|\ll1$, where the ground-state Heitler-London
ansatz is self-consistent. We plot $J(B)$ {[}Eq.~(\ref{J}){]} in
Fig.~\ref{Jplots}(a) and observe the singlet-triplet crossing, where
the sign of $J$ changes from positive to negative (for the parameters
chosen in Fig.~\ref{Jplots}(a) at $B\approx1.3\,{\textrm{T}}$).
Finally, $J$ is suppressed exponentially, $\propto\exp(-2d^{2}b)$,
either by compression of the electron orbitals through large magnetic
fields ($b\gg1$), or by large distances between the dots ($d\gg1$),
where in both cases the orbital overlap of the states in the two dots
is reduced. The Heitler-London result {[}Eq.~(\ref{J}){]} was refined
by taking higher levels and double occupancy of the dots into account
(implemented in a Hund-Mullikan approach), which leads to qualitatively
similar results~\cite{BLD}, in particular concerning the singlet-triplet
crossing. These results have been confirmed by numerical calculations
which take more single-particle levels into account \cite{Hu}. 

A characterization of a double dot can be performed with transport
measurements. We describe transport through a double quantum dot,
using a master equation approach \cite{ddTransportGL}. We calculate
differential conductance $G=dI/dV_{\mathrm{SD}}$ as a function of
the bias voltage $V_{\mathrm{SD}}=\Delta\mu/e$ in the sequential
tunneling and cotunneling regime. We obtain the main peak of the Coulomb
blockade diamond and its satellite peaks. Since the positions of these
peaks are related to the interdot tunnel splitting and to the singlet-triplet
splitting $J$, one can determine these values in a standard transport
experiment. Further, our model can be checked independently, since
we also predict which satellite peaks have positive or negative values
of $G$ and since we describe structures inside the Coulomb blocked
diamonds which are due to a combined effect of cotunneling and sequential
tunneling \cite{ddTransportGL}. When we measure transport properties
of a structure resembling a single dot, we observe features as would
be expected for a double dot \cite{exchangeZumbuhl}. This indicates
that a double dot in formed within our structure. We can then extract
the $B$-dependent exchange coupling from our data which again is
in agreement with theoretical predictions for double dots, see Fig.
\ref{Jplots}(b). That singlet-triplet crossings occur in single dots
is established experimentally \cite{STcrossing}.

In further experiments, we measured a double quantum dot with tunable
tunnel couplings. Spectroscopy of the double dot states was performed
using a quantum point contact (QPC) as a local charge sensor. From
the charge distribution on the double dot, we can deduce charge delocalization
as a function of temperature and strength of tunnel coupling. Conversely,
we can measure the tunnel coupling $t$ as function of the voltage
applied on a gate in the coupling region. We find that the tunneling
coupling is tunable from $t=0$ to $t=22\:\mu\mathrm{eV}$ when the
gate voltage is increased \cite{ddLargeTunnelCM}.

For few-electron quantum dots, the charging energies of a double quantum
dot can be tuned such that there is only a single electron in each
dot. The number of electrons on the dots can be controlled by simultaneously
measuring the charge distribution with a QPC charge sensor \cite{ddSmallLK},
see Fig. \ref{fig:doubleDot}, or by measuring transport through the
double dot \cite{ssSmallRMW}.

\section{Spin relaxation}

\label{secRelax}

The lifetime of an electron spin is described by the following two
time scales. The (longitudinal) spin relaxation time $T_{1}$ describes
the time scale of a spin-flip process when the electron is aligned
along the external magnetic field. The spin decoherence time $T_{2}$
is the lifetime of a coherent superposition $\alpha\spup+\beta\spdown$.
Since quantum gate operations require coherence of the underlying
qubits, they must be carried out on times shorter than $T_{2}$. We
note that $T_{2}\leq2T_{1}$ and typically even $T_{2}\ll T_{1}$
\cite{Abragam}, thus from the sole knowledge of $T_{1}$, no lower
bound for $T_{2}$ can be deduced. Therefore, it is of interest to
investigate the interactions leading to decoherence (as we do now)
and to find ways of measuring the decoherence time $T_{2}$ in an
experiment (see Sec.~\ref{ssecT2}).

For spins on quantum dots, one possible source of spin relaxation
and decoherence is spin-orbit interaction. Calculations show that
phonon-assisted spin-flip times \cite{KhaetskiiNazarov,Erlingsson}
in quantum dots are unusually long. This is so because the spin-orbit
coupling in two-dimensions (2D) is linear in momentum, both for Dresselhaus
and Rashba contributions. Due to this linearity, the effective magnetic
field due to spin-orbit fluctuates transversely to the external magnetic
field (in leading order). This implies that $T_{2}=2T_{1}$ for spin-orbit
interaction \cite{SOT2} and thus long decoherence times are expected.
Another source of decoherence is the hyperfine coupling between electron
spin and nuclear spins in a quantum dot~\cite{BLD,KhaetskiiHf,KhaetskiiHfPRB},
since all naturally occurring Ga and As isotopes have a nuclear spin
$I=3/2$. It is known that such decoherence can be controlled by a
large magnetic field or by polarizing the nuclear spins, i.e., by
creating an Overhauser field~\cite{BLD}.

The spin relaxation time $T_{1}$ of single electron spins on quantum
dots was measured in recent experiments. One way to assess $T_{1}$
is to measure transport through the dot while applying double-step
pulses to the gate voltage of the dot. First, the dot is emptied and
filled again with one electron with a random spin. Then, the electron
is held in the dot during a time $t_{h}$. Finally, the gate voltage
is tuned such that the electron can tunnel out of the dot and contribute
to a current, but only if it is in the excited spin state. Thus, the
(time-averaged) current will proportional to the probability of having
an excited spin on the dot after time $t_{h}$; this probability decays
on the time scale of $T_{1}$. In these experiments, the limited current
sensitivity puts an upper bound on $t_{h}$. Since $T_{1}$ turned
out to be longer than this bound, one was not able to measure $T_{1}$.
Still, it is possible to obtain a lower bound of for $T_{1}$ and
$\approx100\:\mu\mathrm{s}$ was obtained for triplet to singlet transitions
\cite{FujisawaT1} and for $N=1$ Zeeman levels \cite{HansonT1}.
Using a charge read-out device (see Sec. \ref{secSpinMeas}), single
tunneling events can be observed. This allowed us to measure $T_{1}$
directly and $T_{1}^{\mathrm{exp}}=1\:\textrm{ms}$ was obtained at
$B=8\:\mathrm{T}$ \cite{ssReadOutDelft}. We now compare this value
with theoretical predictions \cite{SOT2}. We assume a GaAs dot with
Dresselhaus spin-orbit interaction $H_{\mathrm{SO}}=\beta(-p_{x}\sigma_{x}+p_{y}\sigma_{y})$,
with quantum well thickness $d=5\:\mathrm{nm}$, and with lateral
size quantization energy $\hbar\omega_{0}=1.1\:\mathrm{meV}$, corresponding
to a Bohr radius $a_{\mathrm{B}}=32\:\mathrm{nm}$. The material parameters
are the dielectric constant $\kappa=13.1$, coupling constant of deformation
potential $\Xi_{0}=6.7\:\mathrm{eV}$, piezoelectric constant $h_{14}=-0.16\:\mathrm{C}/\mathrm{m}^{2}$,
sound velocity $s_{j}$ for branch $j$, namely $s_{1}=4.73\times10^{5}\:\mathrm{cm}/\mathrm{s}$
and $s_{2}=s_{3}=3.35\times10^{5}\:\mathrm{cm}/\mathrm{s}$, sample
density $\rho_{c}=5.3\times10^{3}\:\mathrm{kg}/\mathrm{m}^{3}$, and
effective mass $m^{*}=0.067m_{e}$. The remaining unknown parameter
is the spin-orbit length $\lambda_{\mathrm{SO}}=\hbar/m^{*}\beta$.
It can be extracted from (independent) weak antilocalization measurements
\cite{ZumbuhlSO}, where $\lambda_{\mathrm{SO}}\approx9\:\mu\mathrm{m}$
was found. Taking the Zeeman splitting used in the measurement of
$T_{1}^{\mathrm{exp}}$, we obtain \cite{SOT2} $T_{1,\,\mathrm{SO}}^{\mathrm{th}}\approx750\:\mu\mathrm{s}$,
with an error of $50\%$ due to the uncertainty of the value of the
Zeeman splitting. There is some additional uncertainty on the value
of $\lambda_{\mathrm{SO}}$ which depends on electron density and
growth of the sample. For example, we find $\lambda_{\mathrm{SO}}\approx17\:\mu\mathrm{m}$
in other samples \cite{huibersTauPhiDots}, which would indicate a
longer $T_{1}$ time since $T_{1}\propto\lambda_{\mathrm{SO}}^{2}$
\cite{SOT2}. Within these uncertainties we find an agreement between
experiments and theory, $T_{1}^{\mathrm{exp}}\approx T_{1,\,\mathrm{SO}}^{\mathrm{th}}$.
Moreover, the predicted $B$-dependence \cite{SOT2} of $1/T_{1}$
agrees well with the experiment \cite{ssReadOutDelft}, where a plateau
is seen around $B\sim10\:\mathrm{T}$. From this we can conclude that
the spin-phonon mechanism is the dominant source for spin relaxation
(and not hyperfine interaction). Since $T_{2}=2T_{1}$ for spin-orbit
interaction \cite{SOT2} and since there is no difference between
decoherence and relaxation for hyperfine interaction \cite{KhaetskiiHf,KhaetskiiHfPRB},
we can expect spin decoherence times $T_{2}$ to be on the order of
milliseconds.

\section{Spin Filter}

\label{ssecSpinFilter} An important device for spintronics is a spin
filter which selectively transmits electrons with respect to their
spin orientation. For quantum computation with spin qubits, such a
spin filter can be used for initialization and read out, see Sec.
\ref{ssecSpinQubits} and \ref{secSpinMeas}. We proposed to use a
quantum dot attached to in- and outgoing current leads as a spin filter~\cite{Recher}.
The direction of polarization of this spin filter can be tuned electrically
by changing the gate voltage on the quantum dot. We now describe the
operational principle of such a spin filter and present experimental
implementations~\cite{SFsmallDotLK,SFopenDotsCM,SFclosedDotsCM}. 

\begin{figure}
\centerline{\includegraphics[%
  width=120mm]{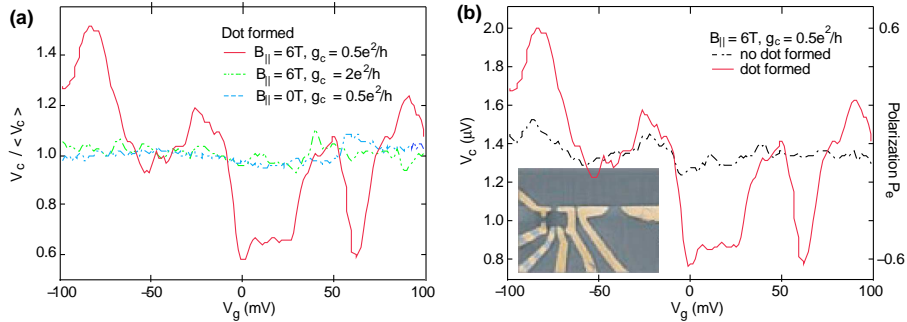}}

\caption{The spin polarization of current through a quantum dot is detected
with an analyzer setup. The polarization is measured via the collector
voltage (at the focusing peak, see text). The polarization of the
current though the quantum dot in a magnetic field fluctuates as function
of gate voltage. The fluctuations in the collector voltage only occur
when the emitter forms a quantum dot, the collector is spin-sensitive,
and an in-plane magnetic field is applied \cite{SFopenDotsCM}. (a)
Comparison of normalized focusing peak height as a function of $V_{g}$
at $B_{\parallel}=6\:\mathrm{T}$ for a spin-selective collector,
$g_{c}=0.5e^{2}/h$ (red curve), at $B_{\parallel}=6\:\mathrm{T}$
for an unpolarized collector, $g_{c}=2e^{2}/h$ (green curve), and
at $B_{\parallel}=0$ with $g_{c}=0.5e^{2}/h$ (blue curve). Dividing
by average peak height, $\expect{V_{c}}$, normalizes for changes
in focusing efficiency. (b) Focusing peak height at $B_{\parallel}=6\:\mathrm{T}$
with spin-selective collector, $g_{c}=0.5e^{2}/h$, comparing an emitter
which is a point contact at $2e^{2}/h$ (black curve) and an emitter
which is a quantum dot with both leads at $2e^{2}/h$ (red curve).
The inset shows a micrograph of the measured device, where the dot
on the left and the QPC on the right side \cite{SFopenDotsCM}. }

\label{fig:filterLarge}
\end{figure}

\begin{figure}
 \centerline{\includegraphics[%
  width=70mm]{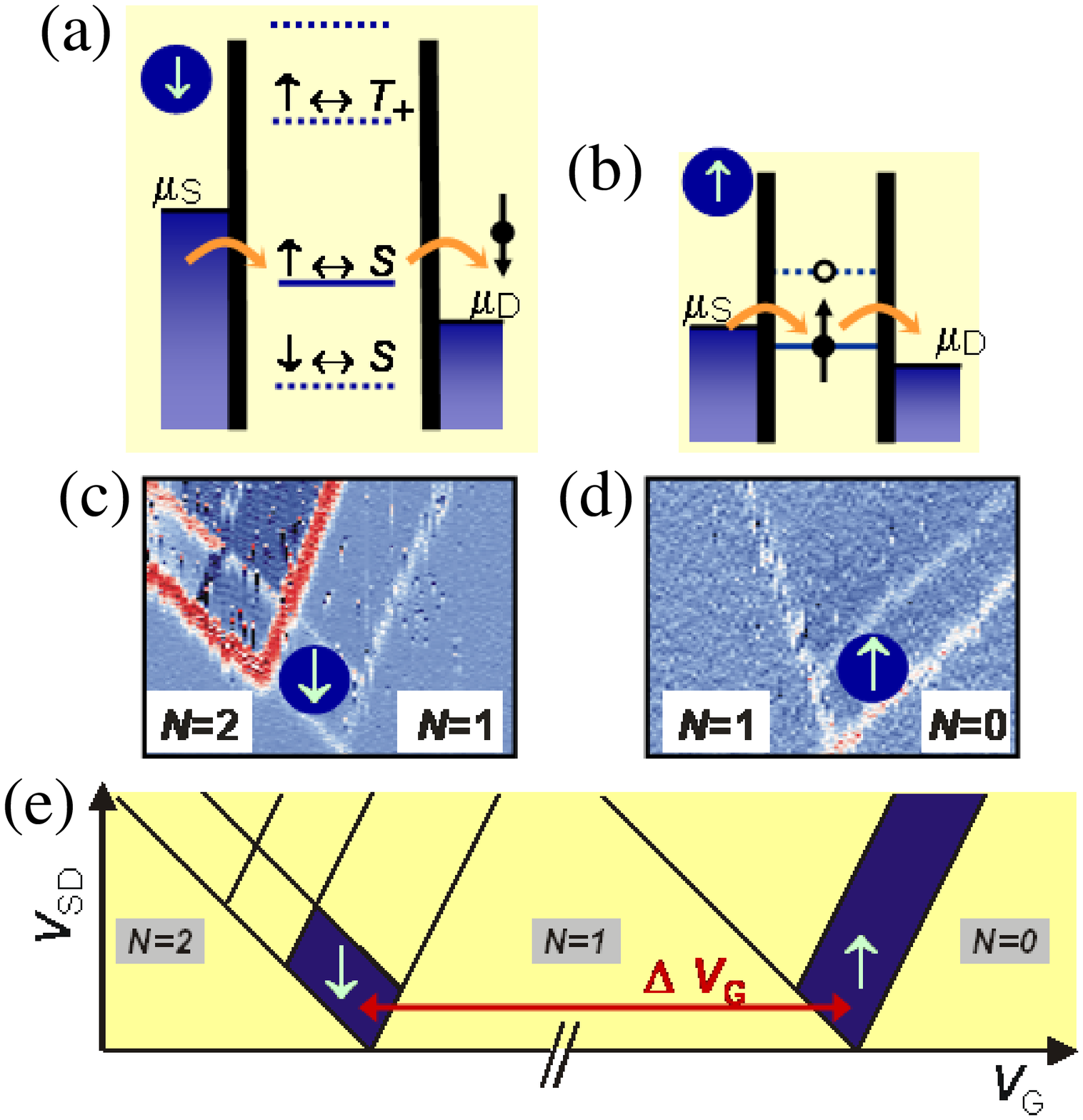}}

\caption{Spin filter in the sequential tunneling regime \cite{Recher}. (a),
(b) Operation principle of the spin filter. (a) Regime where the only
allowed $1\leftrightarrow2$ electron transitions are $\uparrow\leftrightarrow S$
due to energy conservation, thus only spin-$\downarrow$ electron
pass through the dot (see text). (b) The $0\leftrightarrow1$ electron
transition provides a spin filter for spin-$\uparrow$ electrons.
(c), (d) The experimentally measured $dI/dV_{\mathrm{SD}}$ is plotted
as function of bias voltage $V_{\mathrm{SD}}$ and gate voltage $V_{\mathrm{G}}$
at $B_{\parallel}=12\:\mathrm{T}$ \cite{SFsmallDotLK}. In the region
labeled {}``$\downarrow$'' only spin-down electrons pass through
the dot while in the region {}``$\uparrow$'' only spin-up electrons.
(e) Analyzing all transitions between the dot states $\ket{0}$, $\spup$,
$\spdown$, $\ket{S}$, $\ket{T_{0}}$, and $\ket{T_{\pm}}$, the
predicted $dI/dV_{\mathrm{SQ}}$ is shown schematically and agrees
with the experimental data. This indicates that the current is spin
polarized in the regimes labeled by arrows \cite{SFsmallDotLK}. }

\label{fig:spinFilter}
\end{figure}

Our spin filter proposal \cite{Recher} requires a lifted spin-degeneracy
on the dot with a Zeeman splitting $\Delta_{z}=|\mu_{B}gB|$. For
two electrons on the dot, we assume a singlet ground state with energy
$E_{S}$, while the lowest-lying triplet state has a higher energy
$E_{T_{+}}$. Let us consider the sequential tunneling transition
where the number of electrons on the dots changes from 1 to 2. The
bias between the leads at chemical potentials $\mu_{1,\,2}$ is $\Delta\mu=\mu_{1}-\mu_{2}>0$.
For small bias and low temperatures such that $\Delta\mu,\, kT<\mathrm{min}\{\Delta_{z},\, E_{T_{+}}-E_{S}\}$,
only ground state transitions are energetically allowed, i.e., $\spup\leftrightarrow\ket{S}$.
Thus, only spin down electrons can tunnel through the dot, see Fig.
\ref{fig:spinFilter}(a). We calculate the current through the dot
using the standard tunneling Hamiltonian approach in the Coulomb blockade
regime~\cite{Kouwenhoven} and the master equation for the reduced
density matrix of the dot~\cite{Recher}. The current in first order
in tunneling is the sequential tunneling current $I_{s}$~\cite{Kouwenhoven},
which is spin-$\downarrow$ polarized. The second-order contribution
is the cotunneling current $I_{c}$~\cite{averinnazarov} which involves
a virtual intermediate state, where energy conservation can be violated
for a short time. Thus, our energetic argument does not hold here
and the cotunneling current $I_{c}$ contains a spin-$\uparrow$ component,
reducing the efficiency of the spin-filtering effect. For $\Delta_{z}<E_{T_{+}}-E_{S}$,
the ratio of spin-polarized to unpolarized current is~\cite{Recher}\begin{equation}
I_{s}(\downarrow)/I_{c}(\uparrow)\sim\frac{\Delta_{z}^{2}}{(\gamma_{1}+\gamma_{2})\max\{ k_{B}T,\,\Delta\mu\}},\label{efficiencyST}\end{equation}
 where $\gamma_{l}$ is the tunneling rate between lead~$l$ and
the dot. In the sequential tunneling regime we have $\gamma_{l}<k_{B}T,\,\Delta\mu$,
thus, the ratio Eq.~(\ref{efficiencyST}) is large and the spin-filter
is efficient. We implemented this spin filter with a single quantum
dot in the few electron regime \cite{SFsmallDotLK}. The measured
currents agree well with the theoretical predictions, see Fig.~\ref{fig:spinFilter}.

Spin filtering properties of both open \cite{SFopenDotsCM} and
Coulomb blockaded \cite{SFclosedDotsCM} quantum dots were measured
directly in a polarizer-analyzer geometry, where the spin polarization
of current emitted from the dot (polarizer) was detected using a QPC
at $g=0.5e^{2}/h$ (analyzer) \cite{spinAnalyzerPotok}. These polarizer
and analyzer elements were coupled by transverse focusing with the
use of a small magnetic field applied perpendicular to the sample
plane shown in the inset of Fig.~\ref{fig:filterLarge}(b). The collector
voltage at the QPC shows a focusing peak when the distance between
emitter and collector is an integer multiple of the cyclotron diameter.
Measuring at the focusing peak, we find that in the presence of an
in-plane field of a few Tesla or more, the current through the quantum
dot (which is strongly coupled to leads) is indeed spin polarized.
For the case of open dots \cite{SFopenDotsCM}, the direction of polarization
can be readily tuned from along to against the applied in-plane field,
see Fig.~\ref{fig:filterLarge}. However, for the closed dots, reversed
spin filtering was not observed though ground-state peak motion was
seen \cite{SFclosedDotsCM}. More work is needed to clarify this departure
from expectation.

\section{Read-Out of a Single Spin}

\label{secSpinMeas}

At the end of every (quantum) computation, one reads out the result
of the computation. For this it is sufficient to determine the state
of some qubits which are either in state $|\!\uparrow\rangle$ or
in state $|\!\downarrow\rangle$ (we do not need to measure a coherent
superposition). However, it is very hard to detect an electron spin
by directly coupling to its tiny magnetic moment (on the order of
$\mu_{B}$). This difficulty is overcome by converting the spin information
into charge information, which is then measured (we describe implementations
below). Ideally, the qubit state can be determined in a single measurement,
referred to as single shot read out. In general, however, there are
some errors associated with the measurement, thus the preparation and
measurement of the qubit need to be performed not only once but $n$
times. We now determine $n$ by assuming that the measurement has
two possible outcomes, $A_{\uparrow}$ or $A_{\downarrow}$. Then,
for an initial qubit state $\spup,$ with probability $p_{\uparrow}$
the outcome is $A_{\uparrow}$, which we would interpret as {}``qubits
was in state $\spup$.'' However, with probability $1-p_{\uparrow}$,
the outcome is $A_{\downarrow}$ and one might incorrectly conclude
that {}``qubit was in state $\spdown$''. Conversely, the initial
state $\spdown$ leads with probability $p_{\downarrow}$ to $\measOutcome{\downarrow}$
and with $1-p_{\downarrow}$ to $\measOutcome{\uparrow}$. How many
times $n$ do the preparation of a qubit in the \emph{same} initial
state and subsequent measurement need to be performed until the state
of the qubit is known with some given infidelity $\alpha$ ($n$-shot
read out)? We model the read out process with a positive operator
valued measure (POVM) and find from a statistical analysis that we
need~\cite{ELreadout} \begin{eqnarray}
n & \geq & z_{1-\alpha}^{2}\Big(\frac{1}{e}-1\Big),\label{eqnnMin}\\
\efficiencyMeas & = & \left(\sqrt{\binomProb{\uparrow}\binomProb{\downarrow}}-\sqrt{(1-\binomProb{\uparrow})(1-\binomProb{\downarrow})}\right)^{2},\label{eqnMEffG}\end{eqnarray}
 with the quantile (critical value) $z_{1-\alpha}$ of the standard
normal distribution function, $\Phi(z_{1-\alpha})=1-\alpha=\frac{1}{2}\big[1+\mathrm{erf}(z_{1-\alpha}/\sqrt{2})\big]$.
We interpret $\efficiencyMeas$ as \emph{measurement efficiency} \cite{ELreadout},
since it is a single parameter $\efficiencyMeas\in[0,\,1]$ which
tells us if $n$-shot read out is possible. For $p_{\uparrow}=p_{\downarrow}=1$,
the efficiency is maximal, $\efficiencyMeas=100\%$, and single-shot
read out is possible ($n=1$). When the measurement outcome is independent
of the qubit state, i.e., $p_{\downarrow}=1-p_{\uparrow}$ (e.g.,
$p_{\uparrow}=p_{\downarrow}=\frac{1}{2}$), the state of the qubit
cannot be determined and the efficiency is $\efficiencyMeas=0\%$.
For the intermediate regime, $0\%<\efficiencyMeas<100\%$, the state
of the qubit is known after several measurements, with $n$ satisfying
Eq.~(\ref{eqnnMin}). In the more general case, the state of a register
with $k$ different qubits should be determined with infidelity $\beta$.
The probability that the state of all qubits is determined correctly
is $1-\beta=(1-\alpha)^{k}$. One could expect that the required $n$
grows dramatically with $k$. Fortunately this is not the case, from
Eq.~(\ref{eqnnMin}) we find that $n\geq2(1/e-1)\log k/\beta$ is
sufficient.

For the actual implementation of the spin qubit read out, the most
prominent idea is to transfer the qubit information from spin to charge
\cite{Loss97,Recher,ELesr,ELreadout,ssReadOutDelft,KaneSET,FriesenSpinChargeSO},
which can then be accessed experimentally with sensitive voltage or
current measurements. A straightforward concept yielding a potentially
100\% reliable measurement requires a {}``spin-filter'' \cite{Recher}
which allows only, say, spin-up but no spin-down electrons to pass
through, as it is described in Sec.~\ref{ssecSpinFilter}. For performing
a measurement of a spin in a quantum dot, the spin filter is connected
between this dot and a second ({}``reference'') dot. The charge
distribution on this system can be detected with sensitive electrometers
\cite{Devoret} by coupling the dots to a quantum point contact \cite{ddSmallLK,Field}
or to a single-electron transistor (SET) \cite{Rimberg}. Then, if
the spin had been up, it would pass through the spin filter into the
second dot and a change in the charge distribution would be measured,
while there is no change for spin down~\cite{Loss97}. Instead of
a spin filter, one can use different Zeeman splittings on qubit and
reference dot or make use the Pauli principle to read out the spin
qubit via charge detection \cite{ELreadout}.

Finally, we consider the qubit dot coupled to a lead instead of a
reference dot. For Zeeman splittings larger than temperature, one
can tune the dot levels such that only the excited spin state, $\spdown$,
can tunnel into the leads \cite{ELesr} with rate $\gammaOut$ (spin
$\uparrow$ electrons can tunnel only \emph{onto} the dot). Such a
tunneling event changes the number of electrons on the dot and produces
a pulse in the QPC current, whose duration must exceed $\timeQPCmeas$
to be detected, until a spin $\uparrow$ electron tunnels onto the
dot with rate $\gammaIn$. After waiting a time $t$ to detect such
a signal, we have $p_{\uparrow}=1$ and $\efficiencyMeas=p_{\downarrow}=(1-e^{-t\gammaOut})e^{-\timeQPCmeas\gammaIn}$.
We implemented this scheme experimentally \cite{ssReadOutDelft}.
Accounting also for finite $T_{1}$ and temperature, we obtain $p_{\uparrow}=92\%$
and $p_{\downarrow}=70\%$. This means that the measurement efficiency
is $e=41\%$, which is already very close to single-shot read out.
For example, after 16 measurements, one knows the state of a 10 qubit
register with an error smaller than $10^{-4}$. Further, this single
spin detection scheme made it possible to determine the $T_{1}$ times
of electron spins on quantum dots \cite{ssReadOutDelft}, see Sec.~\ref{secRelax}.

\section{Detection of Single-Spin Decoherence}

\label{ssecT2} As it was seen in Sec.~\ref{secRelax}, it is an
important research goal to measure the decoherence time $T_{2}$ of
single spins on quantum dots. For this, we now describe how to extract
the decoherence time $T_{2}$ from the sequential tunneling current
through a quantum dot, in the presence of an applied electron spin
resonance (ESR) field producing spin-flips on the dot \cite{ELesr}.
We assume that the Zeeman splitting on the dot is $g\mu_{B}B>\Delta\mu,k_{B}T$,
while the Zeeman splitting in the leads is different, such that the
effect of the ESR field on the leads can be neglected. This can be
achieved, e.g., by using materials with different $g$-factors for
the dot and the leads. We derive the master equation and find the
stationary reduced density matrix of the quantum dot in the basis
$\spup$, $\spdown$, $\ket{S}$ (with corresponding energies $0=E_{\uparrow}<E_{\downarrow}<E_{S}$).
We can assume that the triplet is higher in energy and does not contribute
to the sequential tunneling current. In the regime $E_{S}>\mu_{1}>E_{S}-g\mu_{B}B>\mu_{2}$,
the current is blocked in the absence of the ESR field due to energy
conservation. We calculate the the stationary current and find~\cite{ELesr}\begin{equation}
I(\omega)\propto\frac{V_{\downarrow\uparrow}}{(\omega-g\mu_{B}B)^{2}+V_{\downarrow\uparrow}^{2}},\label{decoh_curr}\end{equation}
 where the width of the resonance at $\omega=g\mu_{B}B$ is given
by the total spin decoherence rate $V_{\downarrow\uparrow}=(W_{S\uparrow}+W_{S\downarrow})/2+1/T_{2}$.
Here, $W_{S\sigma}$ denotes the rate for the transition from the
state $\ket{\sigma}=\spup,\spdown$ to the singlet $\ket{S}$ due
to electrons tunneling from the leads onto the dot. Therefore, the
inverse of the observed line width $1/V_{\downarrow\uparrow}$ represents
a lower bound for the intrinsic single-spin decoherence time $T_{2}$.
For finite temperatures and in the linear response regime $\Delta\mu<kT$,
the current has roughly the standard sequential tunneling peak shape
$\cosh^{-2}[(E_{S}-E_{\downarrow}-\mu)/2k_{B}T]$ as a function of
the gate voltage $V_{\textrm{gate}}\propto\mu=(\mu_{1}+\mu_{2})/2$,
while the width of the resonance in Eq.~(\ref{decoh_curr}) as a
function of $\omega$ remains unaffected.

\begin{figure}
 \centerline{\includegraphics[%
  height=1.5748in]{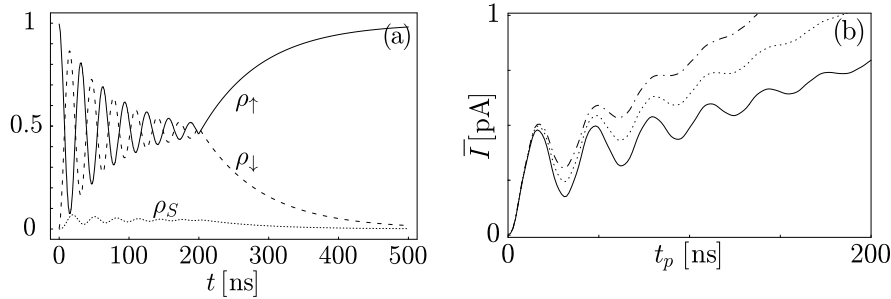}}

\caption{Single spin Rabi oscillations, generated by ESR pulses of length
$t_{p}$, are observable in the time-averaged current $I(t_{p})$
through a quantum dot~\cite{ELesr}. We take the amplitude of ESR
field as $B_{x}^{0}=20\:{\textrm{G}}$ (and $g=2$), and $\Delta\mu>kT$,
$\gamma_{1}=2\times10^{7}\:{\textrm{s}}^{-1}$, $\gamma_{2}=5\gamma_{1}$,
$T_{1}=1\:\mu{\textrm{s}}$, and $T_{2}=150\:{\textrm{ns}}$. (a)
Evolution of the density matrix $\rho$, where a pulse of length $t_{p}=200\:{\textrm{ns}}$
is switched on at $t=0$, obtained via integration of master equation.
(b) Time-averaged current $\bar{I}(t_{p})$ (solid line) for a pulse
repetition time $t_{r}=500\:{\textrm{ns}}$. We also show the current
where $\gamma_{1}$ and $\gamma_{2}$ are increased by a factor of
1.5 (dotted) and 2 (dash-dotted). Calculating the current contributions
analytically, we obtain $\bar{I}(t_{p})\propto\:\:1-\rho_{\uparrow}(t_{p})\,,$
up to a background contribution $\bar{I}_{\textrm{bg}}$ for times
$t<t_{p}$, which is roughly linear in $t_{p}$. Thus, the current
$\bar{I}$ probes the spin state of the dot at time $t_{p}$ and therefore
allows one to measure the Rabi oscillations of a single spin~\cite{ELesr}.}

\label{figRabiOscPulsed}
\end{figure}
 The spin of a quantum dot in the presence of an ESR field shows coherent
Rabi oscillations. It is possible to observe these Rabi oscillations
of a single spin via time-averaged currents when ESR pulses are applied.
Then, the time-averaged current $\bar{I}(t_{p})$ as a function of
the pulse length $t_{p}$ exhibits the Rabi oscillations of the spin-state
of the dot~\cite{ELesr}, see Fig.~\ref{figRabiOscPulsed}. Observing
such Rabi oscillations of a single spin would be a significant achievement,
since this implied an working implementation of a one qubit gate.

\section{Conclusions}

\label{secConclusions}

We described the basic requirements for building a quantum computer
with spin qubits. We addressed several concrete implementation issues
for spin qubits, namely coupling between quantum dots, spin relaxation
and decoherence measurements, spin filter devices, and single-spin
read out setups. For all these issues, we reviewed theoretical and
experimental results. These results give further insight in the details
of quantum computing with spin qubits.

\section{Acknowledgments}

We acknowledge G. Burkard, L. DiCarlo, J.M. Elzerman, J.A. Folk, V.
Golovach, R. Hanson, A. Khaetskii, R. Potok, P. Recher, L.M.K. Vandersypen,
L.H. Willems van Beveren, and D. Zumb\"uhl. This work was supported
by ARO, DARPA, FOM, NCCR Nanoscience, and Swiss NSF.

{}
\end{document}